# Conceptual Insights into the Interaction of the Phase of the Electric Field and Molecular Spectroscopy to Further Understand Molecular Isomerization and Solute-Solvent Interactions

Rachel Glenn

Department of Chemistry, Virginia Tech., Blacksburg, VA, USA and Department of Chemistry, Michigan State University, East Lansing, MI, USA

The real and imaginary parts of the susceptibility are fundamentally associated with the molecular dispersion and absorption, respectively. Measurement of the molecular dispersion has given insight into underlying molecular dynamics such as the identification of conical intersections and solute-solvent interactions. We show that when using a shaped pulse(s), it is possible to mix the real and imaginary parts of the nonlinear susceptibility into the molecular absorption. We then explain how to separate the real and imaginary parts of the nonlinear susceptibility when using a shaped pulse(s). We describe how the imaginary (absorption-like) and real (dispersion-like) parts of the nonlinear susceptibility can be used to understand molecular dynamics involving conical intersections in solvated molecules. We calculate the inhomogenous lineshape for a solvated molecule and show that when the phase is resonant with a vibrational line that the lineshape changes from absorption-like to dispersion-like. We find that the linear polarization is sensitive to the phase of the pulse.

## I. INTRODUCTION

The molecular response is complex, in that, it has both a dispersive (real) and absorptive (imaginary) contribution. Over the course of time, molecular optics has been very fruitful in developing theories based off of experimental observations of the molecular absorption [1–3]. The dispersion spectrum appears to be less interesting than the absorption spectrum, since there are few studies related the dispersion [4–15]. The Kramers-Kronig relation directly relates the absorption and dispersion or the real and imaginary parts of the molecular susceptibility. Thus, the molecular absorption can be used to indirectly measure the dispersion. This is obvious since there is conservation of energy when the molecule and field interact. As a molecular theorist, it is compelling to focus mostly on the theories associated with the molecular absorption, which contains the imaginary part of the susceptibility. The concept of the absorption and dispersion being separately observable, lead us to ask: is it possible for the real part of the susceptibility to be mixed into the molecular absorption signal in the regime of weak electric fields? The molecular absorption signal has both a linear and nonlinear contribution. Certainly, the nonlinear spectrum can be described as a product of real and imaginary parts of the electric-field (multiple interactions with the field) with the real and imaginary parts of the molecular susceptibility. The cross terms between a complex electric field and a complex susceptibility in the nonlinear molecular absorption signal suggest that is possible for an experimentalist to measure an absorption signal that contains the mixing of the real and imaginary parts of the susceptibility. The goal of this paper is to further the conceptual understanding of the complex molecular signal to understand the molecular spectrum and how to utilize it to provide a deeper understanding of the measured or calculated absorption spectrum of molecules.

In the development of this manuscript, we developed an understanding that the absorption spectrum (linear or nonlinear) is very different when a transform limited pulse(s) is employed versus a shaped pulse(s). Here, we consider a shaped pulse as being a complex electric field and a transform limited (TL) pulse as a completely real unshaped electric field. In the third-order nonlinear absorption equations, when a real electric field is used, it is not possible for the real part of the susceptibility to contribute to the absorption signal. When a complex field is applied, the product of a complex field and complex susceptibility mix the real and imaginary parts of the susceptibility into the absorption. The absorption (imaginary part) spectrum will contain a contribution which is dispersion (real part) like. We set out to understand how to isolate the imaginary susceptibility contribution to the nonlinear absorption spectrum and then realized how it can be utilized to further our understanding of molecular processes such as photo-isomerization and solvent-solute interactions.

The separation of the real and imaginary parts of the susceptibility has been employed to understand molecular dynamics such as thermal solvent expansion [7, 11], non-resonant solvent contributions such as the optical Kerr effect [6, 9], and changes in excited-state population [16]. The key to separating the real and imaginary parts of the susceptibility $\chi(\omega)$ has been to control the phase $\varphi(\omega)$ of a reference field [8, 10, 12, 14–16]

$$E(\omega) = A(\omega)M(\omega), \tag{1}$$

where

$$M(\omega) = e^{i\phi(\omega)}, \tag{2}$$

and $A(\omega)$ is the envelope of the field. The measured signal is a product of the field and polarization $P(\omega)\,\alpha\,\chi(\omega)$ [14]

$$\begin{aligned} S_{\pm\phi}(\omega) &= | E_{\pm\phi}(\omega) + iP_{\pm\phi}(\omega) |^2 \\ &\approx | A(\omega) |^2 + A(\omega) \\ &\times (\mathrm{Re}[M_{\pm\phi}(\omega)]\,\mathrm{Im}[P_{\pm\phi}(\omega)] \\ &+ \mathrm{Im}[M_{\pm\phi}(\omega)]\,\mathrm{Re}[P_{\pm\phi}(\omega)]), \end{aligned} \tag{3}$$

where the subscript $\pm\phi$ is used to specify what type of phase is associated with each parameter, i.e. $M_{\pm\phi}(\omega) = e^{\pm i\phi(\omega)}$, $E_{\pm\phi}(\omega) = A(\omega)M_{\pm\phi}(\omega)$ and $P_{\pm\phi}(\omega)$ is expanded as a power series in $E_{\pm\phi}(\omega)$. In this manuscript, we will replace $\pm\phi$ with either $\pm\pi$, $\pm\pi/2$, or TL to specify the phase function as a $\pm\pi$ -phase-step, $\pm\pi/2$ -phase-step, or transform limited (TL-no phase and a pulsed laser beam), respectively. By controlling the phase of the local oscillator to be in phase [ $\phi(\omega) = 0$ ] or quadrature [ $\phi(\omega) = \pi/2$ ], the separation of $\mathrm{Im}[P(\omega)]$ or $\mathrm{Re}[P(\omega)]$ can be obtained. For solvated molecules, the solvent is chosen to be off-resonant with a fairly constant absorption spectrum near the solute absorption peaks. This would suggest that the solute and solvent response do not interact. However, it was shown that the solute and solvent response do interact. Their interaction resulted in different decay profiles for the measured nonlinear absorption and dispersion [13, 16]. The real part of the solute changed sign from positive in the blue to negative in the red portion of the spectrum; whereas, the solvent response remained constant and positive. When the solvent and solute were mixed, the dispersion decay profile was shorten in the red portion of the spectrum and elongated in the blue part of the spectrum [16]. Here we employ an asymmetrically broadened lineshape to model a solvated molecule, by including inhomogenous broadening. A more rigorous approach, which is beyond the scope of this manuscript, would be to include the nuclear degrees of freedom classically [17] or quantum mechanically [18].

Separation of the real and imaginary parts of the signal (hence susceptibility) assisted with identifying the presence of a non-radiative decay pathway through a conical intersection [13]. The non-radiative decay of the wave packet through a conical intersection was to a hot ground state or product state, which occur at different wavelengths with different relaxation rates. Molecules with a fast non-radiative decay, from the excited state to a hot ground-state, have a shift in the ground-state absorption maximum. In turn, this shifts the electronic absorption maximum. Separating the real and imaginary parts of the spectrum, revealed the shift in the real-part (where it crosses from positive to negative). Note that the real-part of the molecular spectrum changes sign due to a $\pi$ -phase change occurring at the absorption maximum. A conical intersection has a phase change associated with the geometrical phase effect. In our previous paper, we described how the geometrical phase effect is observed in a homodyne detected signal. We developed a method to measure a momentum-dependent life-time of a wave-packet through a conical intersection using a single shaped pulse [19]. Here we describe how a single phase-shaped pulse can be used to sense the geometrical phase effect thus identifying the presence of a conical intersection in the heterodyne detected spectrum. At the finalization of this manuscript, we realized its' potential use in the application of controlling chemical reactions. We have included a section describing how this method can utilized for such applications.

It is known that heterodyne detected linear signal is proportional to the modulus squared of the electric field and is not sensitive to the electric field phase. For a solution of molecules the vibrational lines are inhomogeneous broadened. The linear polarization is linearly proportional to the applied field. When the phase of the applied field is in resonance with a vibrational line, it effects the vibrational phase and hence the lineshape. Only particular functions for the electric field phase will have a significant effect on the vibrational lineshape and be observable in the molecular polarization. The phase being able to resonantly excite a particular vibrational line was demonstrated with coherent anti-Stokes Raman scattering (CARS) spectroscopy [20]. When the transition of the $\pm\pi$ -step position coincided with the vibrational resonance, it excited that particular vibrational line. The phase-step creates a sharp change in the electric field at the phase-step transition frequency $\omega_s$ , creating an instantaneous field change from a positive (negative) to negative (positive) value. This impulse when corresponding to an electronic vibrational frequency changes the phase of that particular line and its' inhomogeneous lineshape. The linear response is now sensitive to the applied field phase.

In the following section, we explain how to separate the real and imaginary parts of the susceptibility in a heterodyne detected signal for a single shaped pulse. In Sect. III, we calculate the inhomogeneously broadened vibrational lineshape with a resonant phase. We provide simulations of the molecular spectrum derived in Sect. IV. In Section V, we describe the applications of measuring such a signal to further our understanding of molecular dynamics. In Section VI, we describe how this method can be utilized to control chemical reactions. Concluding remarks are given in Section VII.

## II. THE COMPLEX SUSCEPTIBILITY IN THE HETERODYNE DETECTED ABSORPTION SIGNAL WITH A SHAPED PULSE

Absorption can be described using Fermi's Golden rule, which assumes an exact resonance with the exciting electric field and molecular transition frequency in the long time limit. Fermi's Golden rule, which calculates the transition probability, neglects terms contributing to the lineshape of the molecular signal. For this reason, we do not use Fermi's Golden rule. See Ref. [21] for a more detailed description.

A description of the absorption and dispersion in a molecular signal was developed based upon an assumed type of electric field used to generate the molecular signal [22], specifically continuous wave. To explain how a shaped pulse(s) modifies the previous conceptual ideas of dispersion and absorption, we start with the basic principles of linear and non-linear spectroscopy, and dispersion and absorption. The heterodyne detected signal is given in Eq. 3. The polarization is defined as $P_{\pm\phi}(\omega) = P_{\pm\phi}^{(1)}(\omega) + P_{\pm\phi}^{(3)}(\omega)$, where $P_{\pm\phi}^{(1)}(\omega)$ is the first-order polarization and we do not consider the second-order polarization, since it vanishes for centrosymmetric molecules. The first-order polarization is directly related to the linear susceptibility $\chi^{(1)}(\omega)$ as

$$P_{\pm\phi}^{(1)}(\omega) = E_{\pm\phi}(\omega)\chi^{(1)}(\omega). \tag{4}$$

The third-order polarization is related the third-order susceptibility $\chi^{(3)}(\omega)$ as

$$\begin{aligned} P_{\pm\phi}^{(3)}(\omega) &= \int d\omega_1 d\omega_2 E_{\pm\phi}(\omega_1) E_{\pm\phi}^{*}(\omega_2) \\ &\times E_{\pm\phi}(\omega - \omega_1 + \omega_2)\chi^{(3)}(\omega;\omega_1,\omega_2). \end{aligned} \tag{5}$$

The permittivity $\varepsilon(\omega)$ is commonly defined using a continuous wave laser(s) $\vec{E} = E_0 e^{ik\vec{x} - \omega_L t}\hat{E}$, where $E_0$ is the amplitude of the electric field, $\omega_L$ the frequency of the applied field, and $k$ the wavenumber. From the relation $\vec{D} = \varepsilon_0 \vec{E} + \vec{P} = \varepsilon_0 \varepsilon(\omega)\vec{E}$, it is given as [22]

$$\varepsilon(\omega) = 1 + \chi^{(1)}(\omega) + \chi^{(3)}(\omega)\,|\,\vec{E}\,|^2, \tag{6}$$

where $\mathrm{Re}[\varepsilon(\omega)]$ represents the dispersion and $\mathrm{Im}[\varepsilon(\omega)]$ represents the absorption. If a Gaussian TL pulse(s) is used to generate the molecular signal, the integration in Eq. 5 cannot be done analytically and the factorization of the electric field leading to the relation $\varepsilon(\omega) \propto \chi^{(3)}(\omega)\,|\,\vec{E}\,|^2$ does not occur. However, the Gaussian TL pulse has a frequency independent phase and what is measured is the integration of the envelope of the field with the third-order susceptibility. For a large envelope that is fairly constant within the molecular absorption range, the relation $\varepsilon(\omega) \propto \chi^{(3)}(\omega)\,|\,\vec{E}\,|^2$ approximately holds.

When a shaped pulse is used to generate the molecular polarization, the electric field in the integrand of Eq. 5 has both a real and imaginary contribution that multiplies the complex susceptibility. The nonlinear contribution to the signal, detected in the forward direction, can be described as

$$\begin{aligned}&\mathrm{Im}[E^{*}_{\pm\phi}(\omega)P^{(3)}_{\pm\phi}(\omega)] = \\ &A(\omega)\,\mathrm{Im}[\int d\omega_1 d\omega_2 A(\omega_1)A^{*}(\omega_2)A(\omega-\omega_1+\omega_2) \\ &\times\left\{\Phi_1^{\pm\phi}(\omega,\omega_1,\omega_2)+i\Phi_2^{\pm\phi}(\omega,\omega_1,\omega_2)\right\}\chi^{(3)}(\omega;\omega_1,\omega_2)].\end{aligned} \tag{7}$$

Separation of the phase $M_{\pm\phi}(\omega)M_{\pm\phi}(\omega_1)M_{\pm\phi}(\omega_2)M_{\pm\phi}(\omega-\omega_1+\omega_2)$ into real and imaginary components $\Phi_1^{\pm\phi}(\omega,\omega_1,\omega_2)$ and $\Phi_2^{\pm\phi}(\omega,\omega_1,\omega_2)$ is given in the Appendix A, respectively. The superscript $\phi$ represents the type of phase the functions $\Phi_1^{\pm\phi}$ and $\Phi_2^{\pm\phi}$ have. As described earlier, we will replace $\pm\phi$ with either $\pm\pi$ or $\pm\pi/2$ to represent a $\pm\pi$ or $\pm\pi/2$ phase-step, respectively. The complex field mixes the real and imaginary parts of the susceptibility, into the real and imaginary parts of the permittivity. This raises the question of how to separate the real and imaginary components of the susceptibility, when using a shaped pulse. The separation of the two components of the susceptibility can be accomplished by utilizing the spectral phase of the pulse. Our objective is to understand how a shaped pulse affects the absorption, so any contributions that are not sensitive to the phase will need to be subtracted off. This can easily be done by measuring the molecular signal detected in the forward direction with a single shaped pulse. Afterwards, measuring the same molecular signal with a TL pulse, then subtracting the two, giving

$$\begin{aligned}&S_{TL}(\omega)-S_{\phi}(\omega)\approx A(\omega)\left\{\mathrm{Im}[P_{TL}(\omega)]\right. \\ &\left.-\mathrm{Re}[M_{\phi}(\omega)]\,\mathrm{Im}[P_{\phi}(\omega)]-\mathrm{Im}[M_{\phi}(\omega)]\,\mathrm{Re}[P_{\phi}(\omega)]\right\},\end{aligned} \tag{8}$$

where the subscript $\phi$ (TL) represents the polarization measured with a particular phase (constant phase). When $\mathrm{Re}[M_{\phi}(\omega)]=1$ and $\mathrm{Im}[M_{\phi}(\omega)]=0$, Eq. 8 is zero. When $\mathrm{Im}[M_{\phi}(\omega)]\neq 0$ and $\mathrm{Re}[M_{\phi}(\omega)]\neq 0$, Eq. 8 contains terms proportional to both the real and imaginary parts of $\chi^{(3)}$. When $\mathrm{Im}[M_{\phi}(\omega)]=0$ and $\mathrm{Re}[M_{\phi}(\omega)]\neq 0$, Eq. 8 becomes

$$\begin{aligned}&S_{TL}(\omega)-S_{\phi}(\omega)\approx A(\omega)\int d\omega_1 d\omega_2 A(\omega_1)A^{*}(\omega_2) \\ &\times A(\omega-\omega_1+\omega_2)\{1-\Phi_1^{\phi}(\omega,\omega_1,\omega_2)\} \\ &\times\mathrm{Im}[\chi^{(3)}(\omega;\omega_1,\omega_2)].\end{aligned} \tag{9}$$

Equation 9 represents the change in the absorption spectrum due to a completely real applied spectral phase, which is proportional to the imaginary part of the susceptibility. Another way to measure the imaginary part of the susceptibility is by changing the sign of the applied complex phase. The imaginary part is given by the following relation

$$\begin{aligned}&S_{TL}(\omega)-\tfrac{1}{2}(S_{\phi}(\omega)+S_{-\phi}(\omega))\approx A(\omega)\int d\omega_1 d\omega_2 \\ &\times A(\omega_1)A^{*}(\omega_2)A(\omega-\omega_1+\omega_2)\{1-\Phi_1^{\phi}(\omega,\omega_1,\omega_2)\} \\ &\times\mathrm{Im}[\chi^{(3)}(\omega;\omega_1,\omega_2)].\end{aligned} \tag{10}$$

The real part of the susceptibility can be found by subtracting $S_{\phi}$ and $S_{-\phi}$ giving

$$S_{TL}(\omega) - S_{-\phi}(\omega) \approx A(\omega)\int d\omega_1 d\omega_2$$
$$\times A(\omega_1)A^*(\omega_2)A(\omega-\omega_1+\omega_2)\Phi_2^{\phi}(\omega,\omega_1,\omega_2) \quad (11)$$
$$\times \mathrm{Re}[\chi^{(3)}(\omega;\omega_1,\omega_2)].$$

Equations 9-11 describe how to separate out the real and imaginary components of the susceptibility. In the derivation of Eqs. 9-11, we assumed a non-Kerr-like media and observed both the real and imaginary parts of the nonlinear susceptibility in the nonlinear absorption. For Kerr-like media, mixing of the real and imaginary parts of the nonlinear susceptibility in the nonlinear absorption and dispersion can occur when using TL pulses [23].

The linear signal $\propto E_{\phi}^{\dagger}(\omega)E_{\phi}(\omega)\chi^{(1)}(\omega)$ is insensitive to any spectral phase added to the field and cancels in Eqs. 9-11. $P(\omega)$ should be inhomogenously broadened for solvated molecules,. Typically the phase does not influence an inhomogenously broadened absorption lineshape. However, when the phase is in resonance with a particular vibrational line, it changes the phase of that particular vibrational line and thus influences the absorption lineshape. In the next section, we show how the phase of the electric field influences the absorption lineshape and leads to a non-zero contribution of $\chi^{(1)}(\omega)$ to the detected signals Eqs. 9-11.

### III. THE MOLECULAR LINESHAPE WITH A RESONANTLY APPLIED ELECTRIC PHASE

The influence of the phase of the electric field on the lineshape is most pronounced when the phase-step position coincides with a vibrational excitation. Previous studies considered the phase-step with a sharp transition, such as the Heaviside Theta function, rectangular function, or signum function that has no transition width [24–32]. It has also been considered a function with a continuous transition width, such as the arctangent, error function, or Lorentizian function [20, 33, 34]. Here we approximate the arctangent function with the assumption of a sharp transition. The mathematical derivation of $\pm\pi$ and $\pm\pi/2$ phase-step expressions are given in Appendix C.

The phase-step position is in resonance with an electronic vibrational frequency $\omega_s = \omega_k$, where $\omega_k = (\varepsilon_k - \varepsilon_g)/\hbar$ . The lineshape of an inhomogeneous broadened molecular $P^{(1)}$ with a resonant phase-step transition is given by

$$\left\langle P_{s=k}^{(1)}(\omega) \right\rangle = \int d\omega_k G(\omega_k) E_{\pm\phi}(\omega)\chi_k^{(1)}(\omega)$$
$$= |\mu_k|^2 \, A(\omega) Q_k^{\pm\phi}(\omega,\omega), \quad (12)$$

where the inhomogeneous broadening function $Q_k^{\pm\phi}$ is given by $k$

$$Q_k^{\pm\phi}(\omega_a,\omega_b) = \frac{-1}{2\pi\hbar}\int d\omega_k \frac{\mathcal{G}(\omega_k) M_{\pm\phi}(\omega_a;\omega_k)}{\omega_b - \omega_k + i\gamma_k}. \quad (13)$$

The inhomogenous broadening function $\mathcal{G}(\omega_k)$ is a Gaussian function, defined as

$$\mathcal{G}(\omega_k) = \exp\left[ -\frac{(\omega_k - \langle \omega_k \rangle)^2}{2\Delta_k} \right], \tag{14}$$

where $\langle \omega_k \rangle$ corresponds to the absorption maximum in the measured electronic linear absorption and the full width half max is $2\sqrt{2\ln(2)}\Delta_k$. For simplicity, in Eq. 12, we dropped the $\pm\phi$ notation on $P^{(1)}(\omega)$. It will enter in later when we define the total $P^{(1)}$ signal. The linear absorption signal using Eq. 12 is sensitive to the phase of the pulse. The resonant phase-step changes the phase of the absorption line that is inhomogeneous broadened. In Eq. 12, the absorption-like lineshape is changed to a dispersion-like lineshape. In Fig. 1, we plotted $\mathrm{Im}[\chi^{(1)}(\omega)]$ with the same decay rates to the red and blue of $\omega_k$ (thin-dashed line). The imaginary part of $Q_k$ is plotted in Fig. 1 (thin-solid line) for the same decay rates to the red and blue of $\omega_k$ with a sharp $\pi$-step (Eq. C2), $\tau_s = 2800\mathrm{GHz}^{-1}$. The effect of a resonant phase-step changes the absorption-like lineshape to dispersion-like. When asymmetry is taken into account, using $\gamma_k = 1/50$ fs to the red and $\gamma_k = 1/80$ fs to the blue side of $\omega_k$, the lineshape becomes less dispersive-like. Increasing to $\gamma_k = 1/120$ fs on the blue side, the lineshape is absorption-like.

The first-order contribution to excite all other vibrational lines ($\omega_s \neq \omega_k$) is given as

$$\begin{aligned} \left\langle P^{(1)}_{s\neq k}(\omega) \right\rangle &= \frac{-1}{2\pi\hbar} |\mu_k|^2 A(\omega) M_{\pm\phi}(\omega;\omega_s) \\ &\times \int d\omega_k \frac{\mathcal{G}(\omega_k)}{\omega_b - \omega_k + i\gamma_k}. \end{aligned} \tag{15}$$

The linear polarization generated with a shaped pulse can be considered as a summation of Eqs. 12 and 15

$$\left\langle P^{(1)}_{\pm\phi}(\omega) \right\rangle = \frac{1}{2}\left( \left\langle P^{(1)}_{s=k}(\omega) \right\rangle + \left\langle P^{(1)}_{s\neq k}(\omega) \right\rangle \right). \tag{16}$$

The linear polarization generated with TL shaped pulse is given by

$$\left\langle P^{(1)}_{TL}(\omega) \right\rangle = \left\langle P^{(1)}_{s\neq k}(\omega) \right\rangle \tag{17}$$

with $M(\omega) = 1$ in Eq. 15.

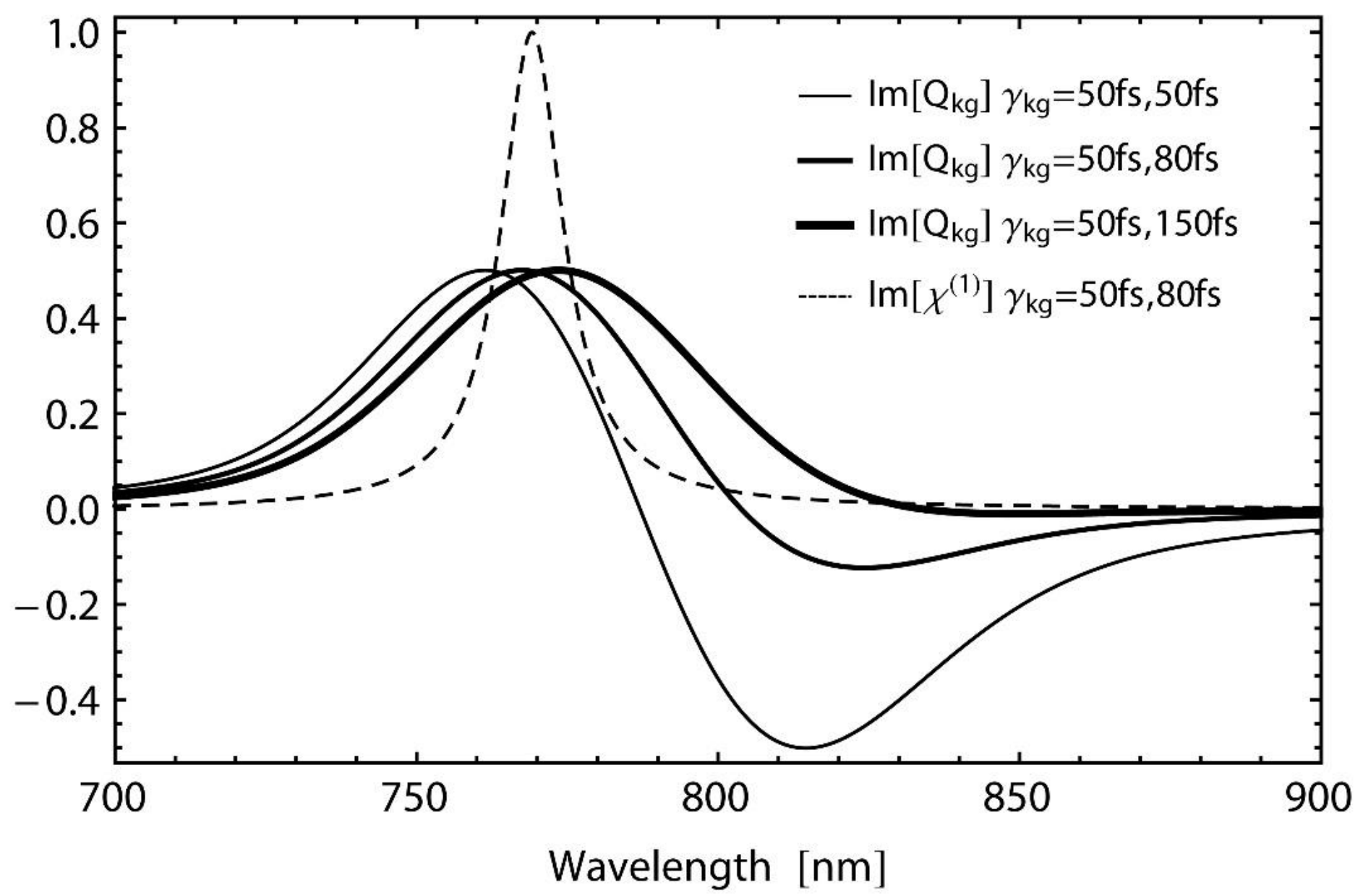

FIG. 1: (Color online) (dashed-black line) The imaginary part of the first-order susceptibility with an symmetric vibrational line width, $\gamma_k = 1/50$ fs and $\lambda_k = 787$nm . ( solid-black) $\mathrm{Im}[Q_k]$ is shown for increasing asymmetry to the blue side, $\gamma_k = 1/50$ fs thin-black, $\gamma_k = 1/80$ fs medium-thin-black, and $\gamma_k = 1/150$ fs thick-black.

The nonlinear polarization, to excite the first electronic state with vibration levels $k$ , will have the susceptibilities $\chi_i^{(3)}$ and $\chi_{iii}^{(3)}$ as given in the Appendix of Ref. [34] in Hilbert space using the loop diagrams, which read

$$\chi_i^{(3)}(\omega,\omega_1,\omega_2) = \left(\frac{-1}{2\pi\hbar}\right)^3 \sum_{kk'} |\mu_{k'}|^2 |\mu_k|^2 \times G_{k'}^{\dagger}(\omega_2) G_g^{\dagger}(\omega_1-\omega) G_k(\omega_2) \tag{18}$$

and

$$\chi_{iii}^{(3)}(\omega,\omega_1,\omega_2) = \left(\frac{-1}{2\pi\hbar}\right)^3 \sum_{kk'} |\mu_{k'}|^2 |\mu_k|^2 \times G_{k'}(\omega) G_g(\omega_1-\omega_2) G_k(\omega_1). \tag{19}$$

The summation over $k$ and $k'$ in Eqs. 18 and 19 includes the possibility of $k \Leftrightarrow k'$ . The inhomogeneous broadening of $P_{k(i)}^{(3)}(\omega)$ will have five different contributions. The first is when $\omega_s = \omega_k$ , and reads

$$
\begin{aligned}
&\left\langle P^{(3)}_{s=k(i)}(\omega)\right\rangle = \left(\frac{-1}{2\pi\hbar}\right)^3 |\mu_{k'}|^2 |\mu_k|^2 \int d\omega_1 d\omega_2 A(\omega_1) \\
&\times A^*(\omega_2) A(\omega-\omega_1+\omega_2) G^\dagger_g(\omega_1-\omega) \int d\omega_k \mathcal{G}(\omega_k) \\
&\times M^*_{\pm\phi}(\omega;\omega_k) M_{\pm\phi}(\omega-\omega_1+\omega_2;\omega_k) M_{\pm\phi}(\omega_1;\omega_k) G_k(\omega_1) \\
&\times \int d\omega_{k'} \mathcal{G}(\omega_{k'}) G^\dagger_{k'}(\omega_2),
\end{aligned}
\tag{20}
$$

where $G_\eta(\omega)$ is the equilibrium Green's function for the state $\eta$

$$
G_\eta(\omega) = \frac{1}{\hbar} \frac{1}{\omega - \omega_\eta + i\gamma_\eta} \tag{21}
$$

$\omega_\eta$ is the vibrational frequency and $\gamma_\eta$ the decay rate. The second contribution is when $\omega_s = \omega_{k'}$ and reads

$$
\begin{aligned}
&\left\langle P^{(3)}_{s=k'(i)}(\omega)\right\rangle = \left(\frac{-1}{2\pi\hbar}\right)^3 |\mu_{k'}|^2 |\mu_k|^2 \int d\omega_1 d\omega_2 A(\omega_1) \\
&\times A^*(\omega_2) A(\omega-\omega_1+\omega_2) G^\dagger_g(\omega_1-\omega) \int d\omega_k \mathcal{G}(\omega_k) \\
&\times M^*_{\pm\phi}(\omega;\omega_k) M_{\pm\phi}(\omega-\omega_1+\omega_2;\omega_k) M_{\pm\phi}(\omega_1;\omega_k) G^\dagger_k(\omega_2) \\
&\times \int d\omega_{k'} \mathcal{G}(\omega_{k'}) G_{k'}(\omega_1),
\end{aligned}
\tag{22}
$$

Third contribution, $\omega_s = \omega_k = \omega_{k'}$ is given by

$$
\begin{aligned}
&\left\langle P^{(3)}_{s=k=k'(i)}(\omega)\right\rangle = \left(\frac{-1}{2\pi\hbar}\right)^3 |\mu_k|^4 \int d\omega_1 d\omega_2 A(\omega_1) \\
&\times A^*(\omega_2) A(\omega-\omega_1+\omega_2) G^\dagger_g(\omega_1-\omega) \int d\omega_k \mathcal{G}(\omega_k) \\
&\times M^*_{\pm\phi}(\omega_2;\omega_k) M_{\pm\phi}(\omega-\omega_1+\omega_2;\omega_k) M_{\pm\phi}(\omega_1;\omega_k) \\
&\times G^\dagger_k(\omega_2) G_k(\omega_1).
\end{aligned}
\tag{23}
$$

The last two are when the phase-step is not in resonance with the vibrational level $k$ or $k'$. This is when $\omega_s \neq \omega_{k'}$, $\omega_s \neq \omega_k$, and $\omega_k = \omega_k$ which reads

$$
\begin{aligned}
&\left\langle P^{(3)}_{s\neq k=k'(i)}(\omega)\right\rangle = \left(\frac{-1}{2\pi\hbar}\right)^3 |\mu_k|^4 \int d\omega_1 d\omega_2 A(\omega_1) \\
&\times A^*(\omega_2) A(\omega-\omega_1+\omega_2) M_{\pm\phi}(\omega-\omega_1+\omega_2;\omega_s) \\
&\times M^*_{\pm\phi}(\omega_2;\omega_s) M_{\pm\phi}(\omega_1;\omega_s) G^\dagger_g(\omega_1-\omega) \int d\omega_k \mathcal{G}(\omega_k) \\
&\times G^\dagger_k(\omega_2) G_k(\omega_1).
\end{aligned}
\tag{24}
$$

The last expressions is when $\omega_k \neq \omega_{k'}$ and $\omega_s \neq \omega_k$ which reads

$$\begin{aligned}
&\left\langle P^{(3)}_{s\neq k\neq k'(i)}(\omega)\right\rangle = \left(\frac{-1}{2\pi\hbar}\right)^3 |\mu_k|^2 |\mu_{k'}|^2 \int d\omega_1 d\omega_2 A(\omega_1) \\
&\times A^*(\omega_2) A(\omega-\omega_1+\omega_2) M_{\pm\phi}(\omega-\omega_1+\omega_2;\omega_s) \\
&\times M^*_{\pm\phi}(\omega_2;\omega_s) M_{\pm\phi}(\omega_1;\omega_s) G^\dagger_g(\omega_1-\omega) \\
&\times \int d\omega_k d\omega_{k'} \mathcal{G}(\omega_k)\mathcal{G}(\omega_{k'}) G^\dagger_k(\omega_2) G_{k'}(\omega_1).
\end{aligned} \tag{25}$$

The inhomogeneous broadening of $P^{(3)}_{(iii)}$, which is related to $\chi^{(3)}_{iii}$ in Eq. 19, will have five contributions, which are given in Appendix B. The inhomogeneous broadened polarization expressions Eqs. 20, 22 - 25 and Eqs. B1- B5, will be used in the next section to calculate the signal detected in the forward direction.

When the lineshape of the electronic vibrational resonances are taken into account, the linear contribution becomes sensitive to the phase of the pulse and no longer vanishes in Eqs. 9-11. Using Eqs. 20, 22 - 25 and Eqs. B1 - B5, the detected third-order polarization generated with a shaped pulse is given as

$$\begin{aligned}
&\left\langle P^{(3)}_{\phi}(\omega)\right\rangle = \tfrac{1}{5}\Big(\left\langle P^{(3)}_{s=k(i)}(\omega)\right\rangle + \left\langle P^{(3)}_{s=k'(i)}(\omega)\right\rangle + \left\langle P^{(3)}_{s=k=k'(i)}(\omega)\right\rangle \\
&+\left\langle P^{(3)}_{s\neq k=k'(i)}(\omega)\right\rangle + \left\langle P^{(3)}_{s\neq k\neq k'(i)}(\omega)\right\rangle + \left\langle P^{(3)}_{s=k(iii)}(\omega)\right\rangle + \left\langle P^{(3)}_{s=k'(iii)}(\omega)\right\rangle \\
&+\left\langle P^{(3)}_{s=k=k'(iii)}(\omega)\right\rangle + \left\langle P^{(3)}_{s\neq k=k'(iii)}(\omega)\right\rangle + \left\langle P^{(3)}_{s\neq k\neq k'(iii)}(\omega)\right\rangle\Big)
\end{aligned} \tag{26}$$

and for a TL pulse, the nonlinear polarization is given as

$$\left\langle P^{(3)}_{TL}(\omega)\right\rangle = \frac{1}{2}\Big(\left\langle P^{(3)}_{s\neq k=k'(i)}(\omega)\right\rangle + \left\langle P^{(3)}_{s\neq k\neq k'(i)}(\omega)\right\rangle + \left\langle P^{(3)}_{s\neq k=k'(iii)}(\omega)\right\rangle + \left\langle P^{(3)}_{s\neq k\neq k'(iii)}(\omega)\right\rangle\Big) \tag{27}$$

with setting $M(\omega)=1$ in Eqs. 24-25 and B4-B5. The factor of 1/5 or 1/2 was added to scale the probability distribution properly. Equation 26 models the nonlinear lineshape with a resonant phase. In the next section, we will simulate the nonlinear heterodyne detected signal with various phase-steps, a $\pm\pi$ -step and $\pm\pi/2$ -step.

## IV. SIMULATIONS

The molecular signal is simulated with a single shaped pulse with phase $\pm\pi$ , $+\pi/2$ , or $-\pi/2$ . We assume that the signal is collected in the forward direction, meaning that the molecular signal is mixed with the laser pulse, which was not absorbed. For this reason, we consider the signal as a single pulse heterodyne detected signal.

All simulations were performed by considering a sharp-spectral phase transition. When using a sharp $\pm\pi$ -step, the imaginary part of the phase contains a sharp Lorentzian, see Fig. 7. When the width of the Lorentzian is smaller than the pulse-shaper resolution, it can be safely neglected, making the $\pm\pi$ -shaped pulse a completely real electric field. The mathematical expressions for the phase-step functions are derived in Appendix C. All calculations presented in this section were calculated using a sharp phase-step $\tau_s = 2800\text{GHz}^{-1}$ and three different phase-step positions $\omega_s$. The laser pulse used was a 13 fs Gaussian pulse at centered at $\lambda_L = 794\text{nm}$, which within the typical range for a Ti-sapphire laser. The inhomogenous broadened function, Eq. 14, had $\langle \lambda_k \rangle = 787\text{nm}$ and $\Delta_k = 0.07$ , which could match a molecule such as ICG in low molar concentration in solution [35]. The symmetric Green's function is accounted for by setting $\gamma_k = \gamma_0 = 1/40\text{fs}$. The asymmetric Green's function was accounted for in Eqs. 20, 22 - 25 by using a frequency dependent decay rate in the Green's function $G_k(\omega)$

$$\gamma_k^{\mp}(\omega_k) = 2\frac{\gamma_0}{(1+0.4*e^{\pm(\omega_k - \langle\omega_k\rangle)})}. \tag{28}$$

When $\omega_k < \langle \omega_k \rangle$, we used $\gamma_k^-$. At the absorption maximum, $\omega_k = \langle \omega_k \rangle$ we set $\gamma_k = \gamma_0$. When $\omega_k > \langle \omega_k \rangle$, we used $\gamma_k^+$ .The ground state level decay rate was set to 100 fs. We chose the electric field strength as $E_0 = 2.5\times10^9\ \text{V/m}$ and dipole moment $\mu = 6D$ to give the most interesting results. The integrations were programmed in C++ and modularized into Python using pybind11 [36]. The one and three dimensional integration were computed with a non-adaptive integration method by specifying the number of sampling points. We tested the integrations using a combination of non-adaptive and adaptive multidimensional quadrature Simpson method [37]. The inner integrations were done using a non-adaptive and the outer integrations done adaptively. We found that the adaptive tolerance was a user chosen control that needed adjustment to ensure enough sampling for the integration to converge correctly. We found this to be difficult when integrating sharp features in a multidimensional integration and resorted to the non-adaptive, slower method, to ensure accurate sampling. The inhomogeneous line-function Eq. 13, appears in some of the multidimensional integrations and was calculated prior to the multi-dimensional integration. The integrated inhomogeneous line-function table was then interpolated using a spline method in the multi-dimensional integrations. All integrations were computed on the high performance computers at the Advanced Research Computing at Virginia Tech.

We first present the subtracted spectrum line-shape

$$\frac{1}{E_0^2}(S_\eta(\omega) - S_{TL}(\omega)) \tag{29}$$

where $\eta$ can be either a $\pm\pi$ , $+\pi/2$ , or $-\pi/2$. We do not neglect the $|P_\eta(\omega)|^2$ contribution to $S_\eta$ or $S_{TL}$. Equation 29 was made dimensionless ($1/E_0^2$ ) and then normalized. Equation 29 is shown in Fig. 2(a-c) for and three different phase-step positions $\lambda_s$ = 777, 787, and 797nm and a $\pm\pi$ -shaped pulse. The absorption signal, Eq. 29, for a TL pulse is also plotted as a reference. The polarization in Eq. 29 is the summation of the first and third-order polarizations, Eqs. 16 (17), 26 (27) for a shaped (TL) pulse. Fig. 2(a) corresponds to the phase-step position below the absorption maximum, Fig. 2(c) above the absorption maximum, and Fig. 2(b) equal to the absorption maximum. As the phase-step crosses the absorption maximum, the sharp feature changes its lineshape from having an asymmetric tail to the right of the absorption maximum to being symmetric and having a tail to the left of the absorption maximum. This is a resonant lineshape effect between the molecular electronic vibrational frequency and phase-step position, as shown in Fig. 1. When the phase-step corresponds to the absorption maximum, the $\pi$-phase from the pulse compensates the phase change at the absorption maximum. What we mean by compensate is that a $\pi$-phase change from the pulse multiplies the $\pi$-phase change at the absorption maximum. This can be understood mathematically as two similar functions: both $\mathrm{Re}[M(\omega)]$ and $\mathrm{Im}[\chi(\omega)]$ have the property of being either positive or negative and changing at the phase-step position or the absorption maximum, respectively. $\mathrm{Im}[M(\omega)]$ and $\mathrm{Re}[\chi(\omega)]$ are both either positive or negative. When the phase-step position and absorption maximum coincide, the multiplication of $\mathrm{Re}[M(\omega)]$ and $\mathrm{Im}[\chi(\omega)]$ changes the molecular dispersion from negative to positive or vise versa. In Figs. 2(a-c), besides the sharp phase-step feature, there is another feature in the spectrum, going negative and positive. We looked at the data and found that the negative contribution was from deconstructive interference occurring between $\mathrm{Im}[E_{\pm\pi}P_{\pm\pi}]$ and $|P_{\pm\pi}|^2$. More specifically from the interference of the $P^{(1)}$ contribution to the signal, which is sensitive to the phase of the pulse. We chose an electric field strength and dipole moment such that the $|P|^2$ contribution is substantial. We checked that when either the dipole moment strength or electric field strength decrease, the signal becomes positive. Inhomogenous broadening of the electronic vibrational lines, causes frequencies to overlap. The overlap of vibrational frequencies with different phase can cause deconstructive/constructive interference.

The signal Eq. 29 is plotted in Fig. 2(d-f) for a $\pi/2$-step shaped pulse. The $M_{\pm\pi}$ and $M_{\pi/2}$ functions both transition from a positive value to a negative value and both have broadened absorption features in Fig. 2. The sharp phase-step feature at 777nm in Fig. 2(d) has an asymmetric lineshape that becomes more symmetric when the phase-step position coincides with the absorption maximum (787 nm) and then again asymmetric when the phase-step is at 797nm. Comparing Figs. 2(a-c) and (d-f), we see that (d-f) do not contain a negative absorption signal. The shaped pulse $\pi/2$ does not create a full $\pi$-phase change of the electronic vibrational frequencies. The signal from a $\pm\pi/2$-step shaped pulse has both a dispersive and absorptive contribution, whereas a $\pm\pi$-step shaped pulses has only an absorptive contribution. To understand the processes occurring, we will need to separate the absorption and dispersion contribution in Fig. 2(d-f).

The signal for a $-\pi/2$-step shaped pulse is shown in Figs. 2(h-j). The signal does not appear to have a broadened absorption lineshape, instead it appears to have a slightly negative absorption. Before with the $\pi$-step, we suggested constructive/deconstructive interference. Deconstructive interference occurs when two signals are out of phase ($\pi$ phase-shift). Using a $-\pi/2$-step, deconstructive interference doesn't occur. Instead, the negative absorption could be viewed as a type of single pulse phase-cycling. There is a phase-change associated with the absorption and emission. The $-\pi/2$ changes the phase of the red spectrum by an additional $\pi/2$, and the blue spectrum by subtracting a $\pi/2$. As previously mentioned, the $-\pi/2$ signal contains both an absorption and dispersion contribution. The only contribution to the signal in Figs. 2(h-j) is the sharp feature that is either symmetric Fig. 2(i) or anti-symmetric Fig. 2(h,j). Comparing the $\pm\pi$ and $\pm\pi/2$ plots, the $-\pi/2$ gives the best lineshape indication of the absorption maximum.

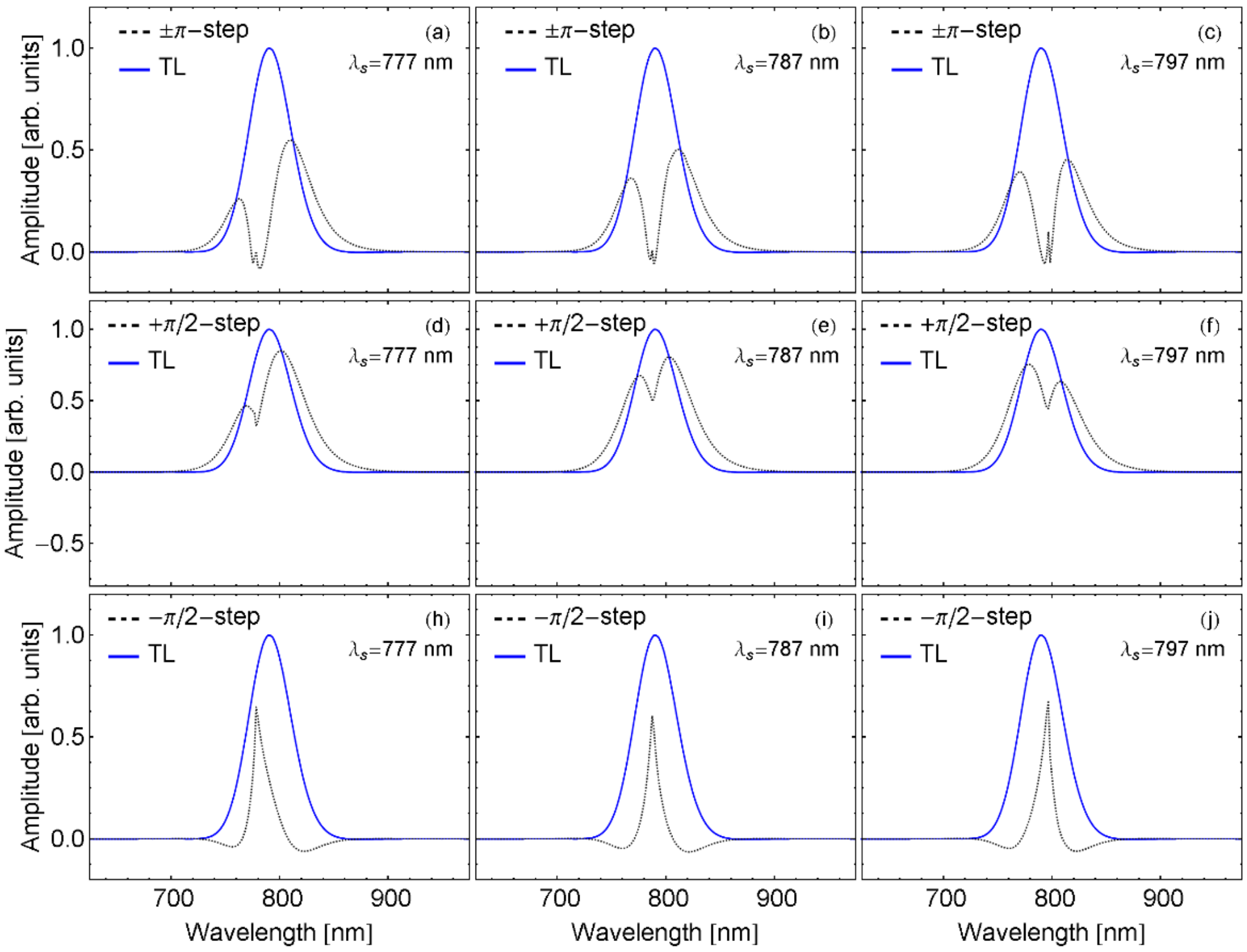

FIG. 2: (Color online) The subtracted spectrum Eq. 29 for three different values of $\lambda_s$, 777, 787, 797nm. (a-c) $\pm\pi$ -step shaped pulse. (d-f) $+\pi/2$-step shaped pulse.(h-j) $-\pi/2$-step shaped pulse. The absorption signal, Eq. 29, for a TL pulse is plotted as a reference in all sub-figures. A symmetric decay rate $\gamma_k = \gamma_0$ was used in the Green's function $G_k$.

To truly understand how the dispersion contributes to the detected signal, we need to separate out the real and imaginary parts. The imaginary part of the susceptibility, which is sensitive to the phase, can be separated by using Eq. 29. This entails detecting the spectrum while using a $\pm\pi$ -step and then using a TL shaped pulse and subtracting them

$$\frac{1}{E_0^2}\left(S_{TL}(\omega) - S_{\pm\pi}(\omega)\right). \tag{30}$$

The field intensity $|E_\eta|^2$ cancels in the subtraction and the contribution from $|P^{(1)}|^2$, which is not sensitive to the phase of the pulse, cancels. Another method to isolate the imaginary part of the susceptibility is to use a completely complex shaped pulse, such as the $\pi/2$-step and Eq. 10, which reads

$$\frac{1}{E_0^2}\left(S_{TL}(\omega) - \tfrac{1}{2}(S_{-\pi/2}(\omega) + S_{+\pi/2}(\omega))\right), \tag{31}$$

Equations 30, 31 are plotted in Fig. 3(a-c) for three different values of $\lambda_s =$ 777, 787, and 797 nm. Comparing Eqs. 30 and 31 plotted in Fig. 3, both are nearly identical. The absorption spectrum has a Gaussian shape as what is typically expected, with a sharp dip near the resonant phase-step function. Suggesting that less was absorbed around the phase-step transition. The sharp dip has the symmetric/asymmetric lineshape that we saw in Fig. 2. One might ask if we see any features here related to dispersion? The sharp feature Fig. 3 (black-thin line) has an asymmetric lineshape that is sensitive to the position of the phase-step relative to the absorption maximum. Typically, the real part of the molecular polarization is sensitive to the change in sign from the absorption maximum; however, we see that the imaginary part shows sensitivity. This is because the resonant phase-step changes the electronic vibrational absorption line-shape to a dispersive lineshape, see Eq. 13 and Fig. 1.

The real part of the susceptibility that is sensitive to the phase of the pulse can be found using Eq. 11 and a $\pm\pi/2$-step

$$\frac{1}{E_0^2}\left(S_{-\pi/2}(\omega) - S_{+\pi/2}(\omega)\right). \tag{32}$$

Equation 32 is plotted in Fig. 4(a-b). The dispersion contribution has a sharp peak near the phase-step. One could describe the shape of the curve as the normal dispersion curve multiplied by a negative number at the inflection point, which stops the signal from going negative. The resonant phase-step changes from a positive to a negative at the inflection point. The product of the resonant phase-step and susceptibility flips the normal dispersion curve at the inflection point.

Recapping, the separated absorption/dispersion signal from Figs. 3 and 4 help explain the features seen in Fig. 2. Interpretation of Fig. 2 without the help of both the absorption and dispersion is difficult and could lead to an incorrect interpretation of the processes that occur during absorption and emission. For example, the signal cannot be so easily described as single pulse phase-cycling [38]. Note that it is very easy to see the asymmetry/symmetry in the lineshape of Fig. 4(a-c), which is symmetric at the phase-step position.

## V. PHASE-STEP AND ITS' USES TO DETECT CONICAL INTERSECTIONS AND SOLUTE-SOLVENT INTERACTIONS

The real and imaginary parts of the susceptibility have enabled us to understand complex molecular dynamics. Previous methods utilized unshaped pulses to separate the real and imaginary parts of the detected spectrum. Here we introduced the concept of utilizing the spectral phase of the pulse to separate out the real and imaginary parts contributing to the absorption spectrum. We showed how the phase effects the lineshape of an electronic vibrational line in the molecular spectrum. The lineshape of the electronic vibrational line is sensitive to the geometric (Berry) phase acquired while propagating and the applied phase from the applied field. See Figs. 2, 3, and 4, the lineshape changes asymmetrically as the phase-step position crosses the absorption maximum (phase change). In our previous paper [19], we showed how the phase of the electric field could affect the angular momentum of the wave packet in the excited state potential energy surface (POS). This led us to develop a method to measure a momentum-dependent lifetime of the wave packet through a conical intersection. A conical intersection is known for its geometrical phase. It is defined as the point where two electronic POSs cross due to nuclear motion. The topology of the two crossing electronic POS is a double cone. When the wave packet transverses around a conical intersection, it acquires a geometric phase. More specifically, the wave packet acquires a $\pi$-phase-shift for a complete circulation or $\pm\pi/2$ for the wave-packet traversing by the conical intersection. When a shaped pulse with an applied phase $\pi$ ($\pm\pi/2$)-step, located at $\omega_s = \omega_k$, excites and creates a wave packet, the wave packet will acquire a $\pi$ ($\pm\pi/2$)-phase shift at $\omega_s = \omega_k$. As the wave packet propagates and encircles a conical intersection, it acquires another $\pi$-phase shift.

We postulate that the applied $\pi$ ($\pm\pi/2$)-phase from the field could cancel with the geometrical (Berry) phase acquired, if that vibrational frequency encounters a conical intersection. Directly changing the phase of the wave packet by changing the phase of the electric field was demonstrated experimentally [39] using two delayed pulses and detecting the florescence. The florescence signal as a function of delay showed a 180 degree phase difference when the two pulses were in phase vs. out of phase (by a factor of $\pi$).

We simulate the wave packet encountering a conical intersection by modifying the Green's function. The Green's function $G_k(\omega)$ represents the propagation of the wave packet in the excited state potential energy surface. We represent the phase acquired by the wave packet intersecting the conical intersection as a sharp $\pi$-phase-step located at 1.59 eV. The sharp phase-step function in the Green's function has a transition frequency of $\tau_s = 2800\text{GHz}^{-1}$ and includes the narrow Lorentzian C1 (Fig. 7b solid red line). The modified Green's function $G_k(\omega)$ in Eqs. 20, 22 - 25 and Eqs. B1 - B5, is given by

$$G_k(\omega) \rightarrow G_k(\omega) M_\pi(\omega), \tag{33}$$

where $M_\pi$ is given by Eq. C1 with $\lambda_s = 777\text{nm}$. To simulate the detected spectrum, we employed a $+\pi/2$ phase-step and plotted the subtracted spectrum Eq. 29 for three different phase-step positions. In Fig. 5c, the position of the phase-step position at 797nm does not coincide with the geometrical phase acquired at 777nm from the conical intersection. There are two dips in Fig. 5c, one from the phase-step location at 797nm and another at 777nm from the simulated geometrical phase change from the wave packet encountering a conical intersection. In Fig. 5b, the phase-step position is changed to 787nm. The line-shape of the phase-step dip at 787nm becomes asymmetric. In Fig. 5a the phase-step location and simulation of the geometrical phase coincide (777nm) and the phase from the electric field and conical intersection combine and change the lineshape to be asymmetric. It is the addition of a $\pi$ (geometrical phase) and $+\pi/2$ (electric field)-phase. Comparing Figs. 2(e) and Fig. 5(b) there is not much change, except for the dip in Fig. 5(b) from the conical intersection. A similar comparison between Figs. 2(f) and Fig. 5(c) can be seen. When comparing Figs. 2(d) and Fig. 5(a), we see as significant difference in the lineshape when the phase-step position coincides with the conical intersection location. This demonstrates that the phase-step lineshape can be utilized to locate where the geometrical phase change occurs due to a conical intersection in the heterodyne detected spectrum.

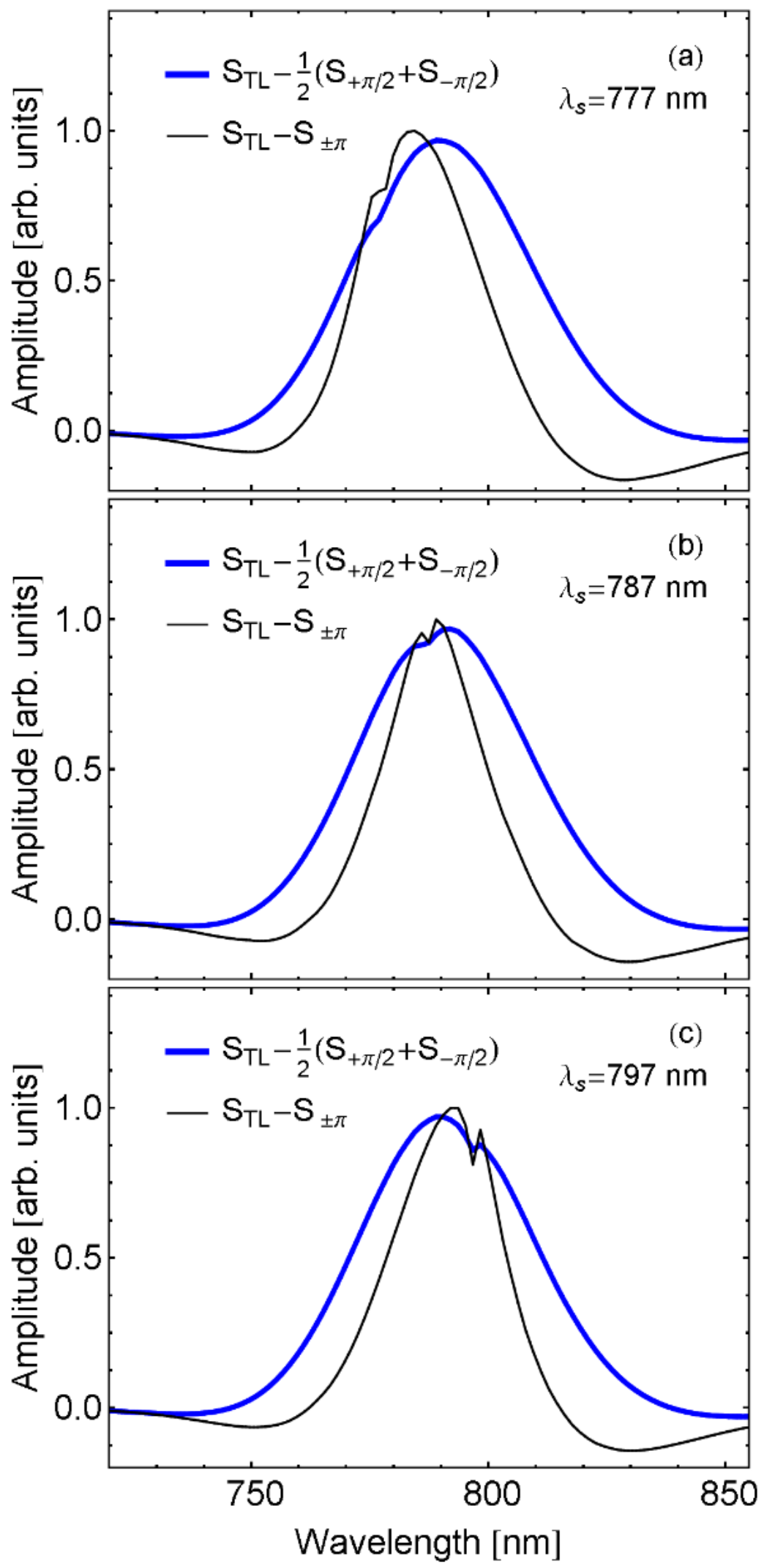

FIG. 3: (Color online) (a-c) The imaginary part of the subtracted spectrum Eq. 30 (thin black line) and 31 (thick blue line) for three different values of $\lambda_s$, 777, 787, 797 nm. A symmetric decay rate $\gamma_k = \gamma_0$ was used in the Green's function $G_k$.

Direct correlation between the electric field phase and wave packet was demonstrated experimentally with using two delayed pulses and detecting the fluorescence from a single molecule [39]. The detected signal showed a $\pi$ -phase difference when the two pulses where out of phase vs. in phase. We showed that a chirped pulse can directly change the angular momentum of the wave packet in the excited stated potential energy surface [19] and be used to control molecular photo-isomerization through deconstructive/constructive interference. At the same time, Ref. [19] was verified experimentally [40]. It was suggested that the inclusion of conservation of angular momentum of the wave packet ensures a correct description of the geometrical phase effect at a conical intersection [41]. Based upon this findings and theory presented here in this manuscript, one can utilize the phase to control the deconstructive/constructive interference occurring at a conical intersection. Controlling the isomer type is an important step in control chemical reactions. The deconstructive/constructive interference can be detected by a change in products [13]. We demonstrated a method to do this theoretically using a chirped pulse in our previous manuscript [19] and experimentally demonstrated it in [40]. Here, we used a simple model to demonstrate that the conical intersection is directly observable in the lineshape of the linear/nonlinear absorption spectrum when using a phase-step. Extending Fig. 5 to a two-dimensional spectrum of the molecular absorption vs. the phase-step position could reveal which vibrational frequencies inherit a $\pm\pi$ or $\pm\pi/2$ geometrical phase acquired as the wave packet transverses around or encircles the conical intersection. The dramatic change in lineshape at a particular phase-step position $\omega_s$, due to the cancellation of the phase as the vibrational energy $\omega_s$ crosses a conical intersection, would be the smoking gun to reveal which electronic vibrational energies encountered the conical intersection. The change in the lineshape is most noticeable when plotting the subtracted signals, Eqs. 30 - 32.

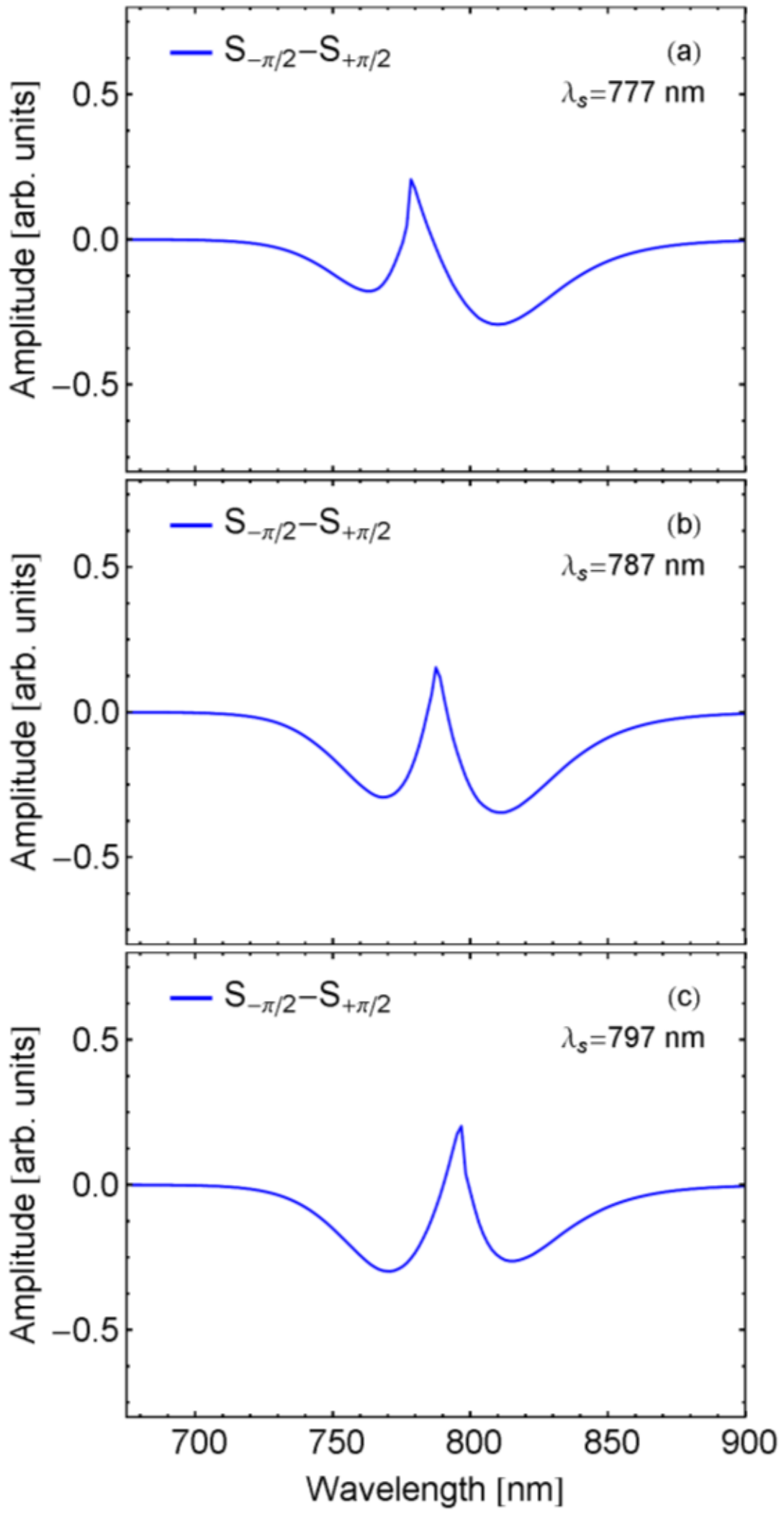

FIG. 4: (Color online) (a-c) The imaginary part of the subtracted spectrum Eq. 32 for three different values of $\lambda_s$, 777, 787, 797 nm. A symmetric decay rate $\lambda_s = \lambda_0$ was used in the Green's function $G_k$ .

We saw how a resonant phase-step was sensitive to the symmetric lineshapes when inhomogeneous broadened. See Fig. 1. Asymmetric lineshapes have been observed in polar [42, 43], nonpolar [44], and non-interacting solvents [45]. Both symmetric and asymmetric lineshapes have been observed in the Raman spectroscopy of Stilbene [46]. In polar solvents, the interaction is mostly Coulombic. The first solvation shell changes the ground state charge distribution and hence equilibrium geometry [47]. Due to the change in molecular charge distribution and geometrical change, the POS curvature changes [47], thus the vibrational phase changes. Depending upon the amount of phase change, a vibrational lineshape may become asymmetric [46]. In fact, vibrational modes that were once forbidden, can be excited in solution. By changing the solvent from a polar to nonpolar solvent any asymmetry from the charge redistribution effect on the lineshape could be identified. In Fig. 1, we saw how pronounced the symmetric vibrational lineshape was when it is resonant with a $\pi$-step. When considering an asymmetric lineshape, Eq. 28, the absorption lineshape changes. We simulate the change of the absorption maximum ( from a hot ground state) with an asymmetric vibrational lineshape Eq. 28 in Fig. 6 for the two different values of the absorption maximum, at 785nm (a,c,e) and 790nm (b,d,f). When the phase-step position is below (lower in wavelength) the absorption maximum in Figs. 6(a, b) the line profile is asymmetric with a longer tail to the longer wavelengths. Changing the phase-step position to be above the absorption maximum Fig. 6(e, f) the asymmetric lineshape changes its shape to have a tail that is longer to the shorter wavelengths. Comparing Fig. 6(c,d), we see that when the absorption maximum changes from 790nm to 785nm that the lineshape changes from a symmetric like to asymmetric like. This demonstrates that the phase-step could be used to detect a hot ground state from a conical intersection or a change in the variation of the peaks due to a viscosity change.

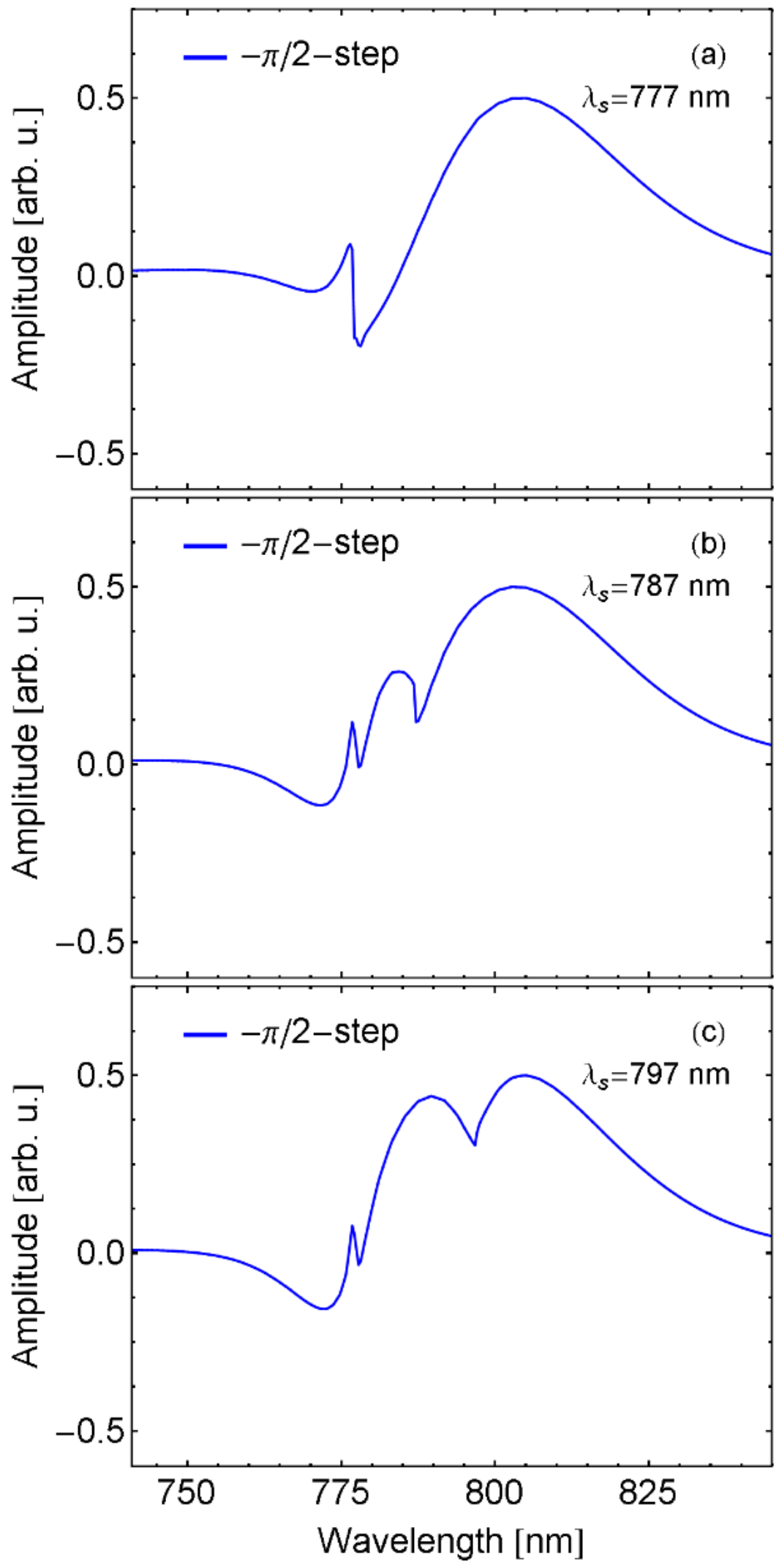

FIG. 5: (Color online) (a-c) The imaginary part of the subtracted spectrum Eq. 29 for three different values of $\lambda_s$, 777, 787, 797 nm. A symmetric decay rate $\lambda_s = \lambda_0$ was used in the Green's function $G_k$ (Eq. 33).

Extending Fig. 6 to a two-dimensional plot of the spectrum vs. the phase-step position for various types of solutes could reveal a more precise change in the asymmetric lineshape. More precisely, comparing the two-dimensional plot of the spectrum vs. the phase-step position, between the two polar/non-polar solvents for Stilbene. Changing the solvent viscosity can slow down the molecular isomerization process and decrease the product yield. This can in turn tell an experimentalist if the observed change in a two-dimensional plot of the phase-step position vs. spectrum is indeed associated with the geometrical phase change due to the conical intersection.

## VI. APPLICATION TO CONTROLLING CHEMICAL REACTIONS

Initially, pulse-shaping was seen a method to enhance or suppress quantum pathways, where the description of the molecule was represented as level scheme. Many fruitful theoretical ideas developed from this assumption, such as control of two-photon absorption, Raman resonances, etc. [48, 49]. Similar to a level-scheme, the photo-isomerization of a molecule has been represented as a two-dimensional potential line with one-to-two different pathways and the possibility of transitioning to other nearby POSs. Three dimensional models of the excited state POS have also been developed. The initial wave packet created by absorption of the electric field in the excited state is typically described as being completely real and independent of the electric field. Gaussian-like wave packets are often chosen. Theories developed from a three-dimensional POS lead to descriptions that could be utilized to further our understanding to control chemical reactions [18, 50, 51]. A different approach to advance our understanding of conical intersections is using genetic algorithms to find an optimal electric field [52] to control the isomerization. To compare these advances in the theoretical description of the wave packet during isomerization in the excited state we need an experimental modality to test the theory. We need to determine the lifetime of the wave packet and what path it propagated along in the excited state POS. To our knowledge, no one has yet to develop an experimental method to map out all of the pathways of the wave packet and its' properties: How sharp is the curvature of the POS; what is the momentum-dependent lifetime of the wave packet through a conical intersection; which electronic vibrational frequencies of the wave packet encircle the conical intersection, transverse by the conical intersection, or do not encounter the conical intersection?

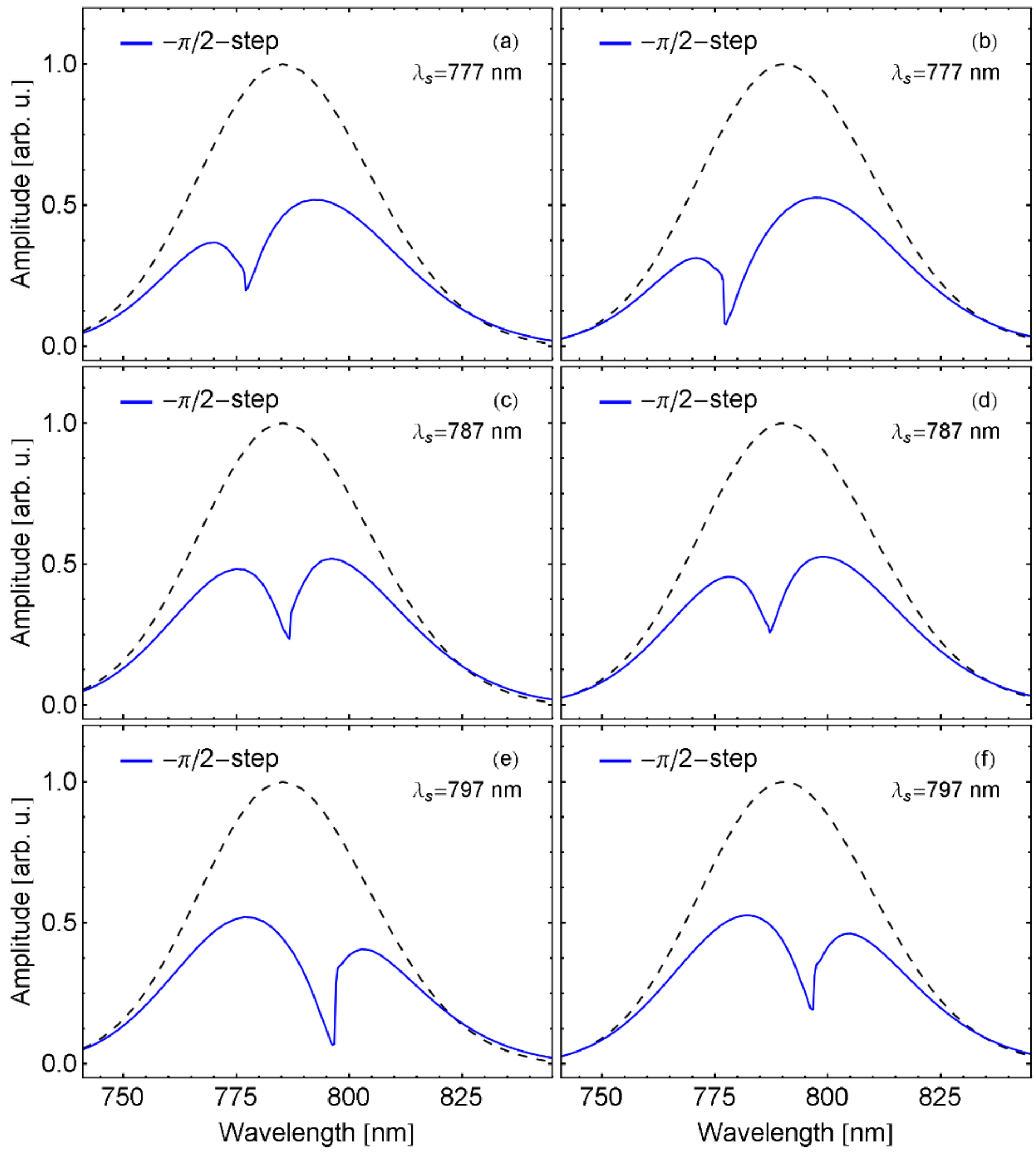

FIG. 6: (Color online) The imaginary part of the subtracted spectrum Eqs. 29 for three different values of $\lambda_s$ , 777, 787, 797 nm. An asymmetric decay rate, Eq. 28, was used in the Green's function $G_k$ . The absorption maximum was set to 785nm (a,c,e) and 790nm (b,d,f).

One can map out the excited-state POS using the wave packet to answer these questions. The wave packet in the excited state POS is energy that is initially created using a laser pulse. It is a probe to create a map of the excited state POS. As it propagates in the excited state, it can take different pathways, all of which have a different geometrical phase acquired. When photo-isomerization occurs, there are pathways having a geometrical phase of $\pi$ , or $\pm\pi/2$ . By mapping out the POS, one could determine the most practical laser pulse: spectral phase, power, and bandwidth to control the wave packet; thus, leading to control of chemical reactions.

A modality that we suggest to map out the excited state POS is based upon the work in this paper and our previous work [19]. It can be very easily implemented, without using multiple pulses or using genetic algorithms. All that is needed is a laser, pulse shaper, spectrometer (heterodyne-detected), and detector (fluorescence).

The fluorescence can be used to decode how steep the potential energy surface is, determine the momentum dependent lifetime of the wave packet though the conical intersection, and provide an indication of deconstructive/constructive interference occurring at the conical intersection. In Ref. [19] we show how the momentum-dependent lifetime of the wave packet can be measured using a chirped pulse and measuring the florescence. The lifetime is described as being from the moment the wave packet is created and then propagates to the minimum of the POS. This includes the possibility of the wave packet encountering a conical intersection. If the wave packet encountered a conical intersection, then deconstructive/constructive interference from the wave packet encircling the conical intersection could occur. Deconstructive interference leaves a unique shape in the fluorescence vs. chirp rate curve. See Ref. [19] for further details.

This method works best for molecules that have a fluorescence state associated with the excited state isomerization. A broadband pulse would be the most ideal to give a good spectral coverage of the excited state POS. When performing these experiments with a shaped single pulse, the power of the laser should be adjusted and chosen with much care and consideration. A $\pi$-step phase, changes the electric field quickly from positive-to-negative in the frequency domain and could lead to a nonlinear index of refraction, if the power is too strong. To determine if a nonlinear index of refraction is occurring, one could perform the experiment with only the solvent and examine the spectrum. It may be that a small effect of a non-linear refractive index is unavoidable in the experiment.

If one finds no experimental evidence of a conical intersection when performing this modality, then it could be suggested that the bandwidth of the laser and central frequency may need further consideration or that the molecule is not isomerizing. Another option would be to employ multiple pulses with different carrier frequencies to position the wave packet correctly in the POS. Then utilize the phase which is described here and in Ref. [19] to map the POS.

The theory to control isomerization, using only-spectral phase to control the wave packet in the excited POS, was developed in our previous paper [19]. Once the location of the conical intersection in spectrum is known, one can choose the bandwidth to be higher or lower in energy than the location of the conical intersection. This would lead to a higher percentage of the cis-isomer vs. trans-isomer or vice versa.  Pulse shaping can also be employed to influence the wave packet's path in the excited state POS driving the molecule to the desired state and controlling the chemical reaction.

## VII. CONCLUSION

Simulation of molecular isomerization using shaped pulses and quantum chemistry calculations would require employing a wave packet that has both real and imaginary parts. We are unaware of any quantum chemistry code available to accurately calculate the isomerization process with including both propagation on the ground and excited state potential energy surfaces and considering a complex wave packet. Here, we carefully selected a very good example to demonstrate the quantitative ability of the theory to detect a geometrical phase change of an intrinsic property of the molecule: The peak in the absorption maximum and the geometrical phase acquired as the wave packet encircles the conical intersection. We showed how the phase-step can be used to detect the conical intersection location in the spectrum. When a molecule isomerizes, the peak in the absorption maximum shifts. A hot ground state is a good example of this [13]. When a molecule is changed to a higher viscosity solvent, the peak in the absorption maximum changes. We demonstrated that our theory can detect the location of the absorption maximum. We extended this idea qualitatively to detect the geometrical phase change acquired at the conical intersection.

We also introduced a new conceptual idea of choosing a particular parameter of the phase-function to be in resonance with the electronic absorption line. To our knowledge, this is a new conceptual idea to control individual Raman resonance and study their absorption and dispersion.

One might ask how the phase-step could be used in other applications, such as population inversion [53, 54] (negative absorption) or negative index of refraction materials for controlling the group velocity. These are interesting applications that may have potential and are beyond the scope of this manuscript.

## VIII. ACKNOWLEDGMENTS

Patent Pending. R. G. would like to thank Daniel Crawford for his support to complete this manuscript. The authors acknowledge Advanced Research Computing at Virginia Tech for providing computational resources and technical support that have contributed to the results reported within this paper. URL: http://www.arc.vt.edu. R. G. would like to thank M. E. Raikh and C. Boehme for previous collaborations in projects that helped with my initial ideas that lead to developing this theory [55]. R. G. would also like to thank M. E. Raikh for fruitful discussions. R. G. would also like to acknowledge previous experience with experimental work at Michigan State University with M. Dantus that lead to some of the ideas in this manuscript. I am very grateful for my time with M. Dantus.

## Appendix A: $\Phi_1$ and $\Phi_2$ in Eq. 7

The expressions for $\Phi_1$ and $\Phi_2$ in Eq. 7 are given by

$$
\begin{aligned}
&\Phi_1^{\pm\phi}\left(\omega,\omega_1,\omega_2\right)= \\
&-\mathrm{Im}[M\left(\omega\right)]\,\mathrm{Im}[M\left(\omega\right)]\,\mathrm{Im}[M\left(\omega\right)]\,\mathrm{Im}[M(\omega-\omega_1+\omega_2)] \\
&+\mathrm{Re}[M\left(\omega\right)]\,\mathrm{Re}[M\left(\omega\right)]\,\mathrm{Im}[M\left(\omega_2\right)]\,\mathrm{Im}[M(\omega-\omega_1+\omega_2)] \\
&-\mathrm{Re}[M\left(\omega\right)]\,\mathrm{Im}[M\left(\omega_1\right)]\,\mathrm{Re}[M\left(\omega_2\right)]\,\mathrm{Im}[M(\omega-\omega_1+\omega_2)] \\
&+\mathrm{Re}[M\left(\omega\right)]\,\mathrm{Im}[M\left(\omega_1\right)]\,\mathrm{Im}[M\left(\omega_2\right)]\,\mathrm{Re}[M(\omega-\omega_1+\omega_2)] \\
&+\mathrm{Im}[M\left(\omega\right)]\,\mathrm{Re}[M\left(\omega_1\right)]\,\mathrm{Re}[M\left(\omega_2\right)]\,\mathrm{Im}[M(\omega-\omega_1+\omega_2)] \\
&-\mathrm{Im}[M\left(\omega\right)]\,\mathrm{Re}[M\left(\omega_1\right)]\,\mathrm{Im}[M\left(\omega_2\right)]\,\mathrm{Re}[M(\omega-\omega_1+\omega_2)] \\
&+\mathrm{Im}[M\left(\omega\right)]\,\mathrm{Im}[M\left(\omega_1\right)]\,\mathrm{Re}\left[M\left(\omega_2\right)\right]\,\mathrm{Re}[M(\omega-\omega_1+\omega_2)] \\
&+\mathrm{Re}[M\left(\omega\right)]\,\mathrm{Re}[M\left(\omega_1\right)]\,\mathrm{Re}[M\left(\omega_2\right)]\,\mathrm{Re}[M(\omega-\omega_1+\omega_2)]
\end{aligned}
\tag{A1}
$$

$$
\begin{aligned}
&\Phi_2^{\pm\phi}\left(\omega,\omega_1,\omega_2\right)= \\
&-\mathrm{Re}[M\left(\omega\right)]\,\mathrm{Im}[M\left(\omega\right)]\,\mathrm{Im}[M\left(\omega\right)]\,\mathrm{Im}[M(\omega-\omega_1+\omega_2)] \\
&+\mathrm{Im}[M\left(\omega\right)]\,\mathrm{Re}[M\left(\omega_1\right)]\,\mathrm{Im}[M\left(\omega_2\right)]\,\mathrm{Im}[M(\omega-\omega_1+\omega_2)] \\
&-\mathrm{Im}[M\left(\omega\right)]\,\mathrm{Im}[M\left(\omega_1\right)]\,\mathrm{Re}[M\left(\omega_2\right)]\,\mathrm{Im}[M(\omega-\omega_1+\omega_2)] \\
&+\mathrm{Im}[M\left(\omega\right)]\,\mathrm{Im}[M\left(\omega_1\right)]\,\mathrm{Im}[M\left(\omega_2\right)]\,\mathrm{Re}[M(\omega-\omega_1+\omega_2)] \\
&+\mathrm{Im}[M\left(\omega\right)]\,\mathrm{Re}[M\left(\omega_1\right)]\,\mathrm{Re}[M\left(\omega_2\right)]\,\mathrm{Re}[M(\omega-\omega_1+\omega_2)] \\
&-\mathrm{Re}[M\left(\omega\right)]\,\mathrm{Im}[M\left(\omega_1\right)]\,\mathrm{Re}[M\left(\omega_2\right)]\,\mathrm{Re}[M(\omega-\omega_1+\omega_2)] \\
&+\mathrm{Re}[M\left(\omega\right)]\,\mathrm{Re}[M\left(\omega_1\right)]\,\mathrm{Im}[M\left(\omega_2\right)]\,\mathrm{Re}[M(\omega-\omega_1+\omega_2)] \\
&-\mathrm{Re}[M\left(\omega\right)]\,\mathrm{Re}[M\left(\omega_1\right)]\,\mathrm{Re}[M\left(\omega_2\right)]\,\mathrm{Im}[M(\omega-\omega_1+\omega_2)]
\end{aligned}
\tag{A2}
$$

where we every $M(\omega)$ term in equations A1 and A2 should have the subscript $\pm\phi$, which we left out in these expressions for ease of reading. They are found by expanding the product of the electric field phase components in the integrand of the third-order polarization.

## Appendix B: inhomogeneous broadened $P_{(iii)}^{(3)}$

There are five contributions to the signal from the inhomogeneous broadening of $P_{(iii)}^{(3)}$. The first is for $\omega_s = \omega_k \neq \omega_{k'}$ and is given by

$$\left\langle P^{(3)}_{s=k=k'(i)}(\omega)\right\rangle = \left(\frac{-1}{2\pi\hbar}\right)^3 |\mu_k|^2|\mu_{k'}|^2 \int d\omega_1 d\omega_2 A(\omega_1)$$
$$A^*(\omega_2)A(\omega-\omega_1+\omega_2)G_g(\omega_1-\omega_2)\int d\omega_k \mathcal{G}(\omega_k)$$
$$M^*_{\pm\phi}(\omega_2;\omega_k)M_{\pm\phi}(\omega-\omega_1+\omega_2;\omega_k)M_{\pm\phi}(\omega_1;\omega_k) \tag{B1}$$
$$G_k(\omega_1)\int d\omega_{k'}\mathcal{G}(\omega_{k'})G_{k'}(\omega)$$

The second contribution corresponds to $\omega_s = \omega_{k'} \neq \omega_k$ and is given by

$$\left\langle P^{(3)}_{s=k'(iii)}(\omega)\right\rangle = \left(\frac{-1}{2\pi\hbar}\right)^3 |\mu_k|^2|\mu_{k'}|^2 \int d\omega_1 d\omega_2 A(\omega_1)$$
$$A^*(\omega_2)A(\omega-\omega_1+\omega_2)G_g(\omega_1-\omega_2)\int d\omega_{k'} \mathcal{G}(\omega_{k'})$$
$$M^*_{\pm\phi}(\omega_2;\omega_{k'})M_{\pm\phi}(\omega-\omega_1+\omega_2;\omega_{k'})M_{\pm\phi}(\omega_1;\omega_{k'}) \tag{B2}$$
$$G_{k'}(\omega)\int d\omega_k\mathcal{G}(\omega_k)G_k(\omega_1)$$

and the third contribution is when $\omega_s = \omega_k = \omega_{k'}$ and reads

$$\left\langle P^{(3)}_{s=k'=k(iii)}(\omega)\right\rangle = \left(\frac{-1}{2\pi\hbar}\right)^3 |\mu_k|^4 \int d\omega_1 d\omega_2 A(\omega_1)A^*(\omega_2)$$
$$A(\omega-\omega_1+\omega_2)G_g(\omega_1-\omega_2)\int d\omega_k \mathcal{G}(\omega_k)M^*_{\pm\phi}(\omega_2;\omega_k) \tag{B3}$$
$$M_{\pm\phi}(\omega-\omega_1+\omega_2;\omega_k)M_{\pm\phi}(\omega_1;\omega_k)G_k(\omega)G_k(\omega_1)$$

There are two other contributions when the phase-step is not in resonance with the vibrational line. The first being $\omega_s \neq \omega_{k'} = \omega_k$

$$\left\langle P^{(3)}_{s\neq k=k'(iii)}(\omega)\right\rangle = \left(\frac{-1}{2\pi\hbar}\right)^3 |\mu_k|^4 \int d\omega_1 d\omega_2 A(\omega_1)A^*(\omega_2)$$
$$A(\omega-\omega_1+\omega_2)M_{\pm\phi}(\omega-\omega_1+\omega_2;\omega_s)M_{\pm\phi}(\omega_1;\omega_s) \tag{B4}$$
$$M^*_{\pm\phi}(\omega_2;\omega_s)G_g(\omega_1-\omega_2)\int d\omega_k \mathcal{G}(\omega_k)G_k(\omega)G_k(\omega_1)$$

and the second one $\omega_s \neq \omega_{k'} \neq \omega_k$ and $\omega_s \neq \omega_k$ reads

$$\left\langle P^{(3)}_{s\neq k\neq k'(iii)}(\omega)\right\rangle = \left(\frac{-1}{2\pi\hbar}\right)^3 |\mu_k|^2|\mu_{k'}|^2 \int d\omega_1 d\omega_2 A(\omega_1)A^*(\omega_2)$$
$$A(\omega-\omega_1+\omega_2)M_{\pm\phi}(\omega-\omega_1+\omega_2;\omega_s)M_{\pm\phi}(\omega_1;\omega_s) \tag{B5}$$
$$M^*_{\pm\phi}(\omega_2;\omega_s)G_{g_1}(\omega_1-\omega_2)\int d\omega_{k'}\omega_k \mathcal{G}(\omega_{k'})\mathcal{G}(\omega_k)$$
$$G_k(\omega)G_{k'}(\omega_1)$$

**Appendix C: Phase-step Function**

We consider a $\pm\pi$ -step or a $\pm\pi/2$-step for the spectral phase that transitions sharply and has a finite transition width. We use the arctangent function for a $\pm\pi$ -step, dashed-line shown in Fig. 7a. The function $M(\omega)$ reads

$$M_{\pm\pi}(\omega) = \cos[\pi/2 - \arctan(\omega - \omega_s)\tau_s] \pm i\sin[\pi/2 - \arctan[(\omega - \omega_s)\tau_s]]. \quad \text{(C1)}$$

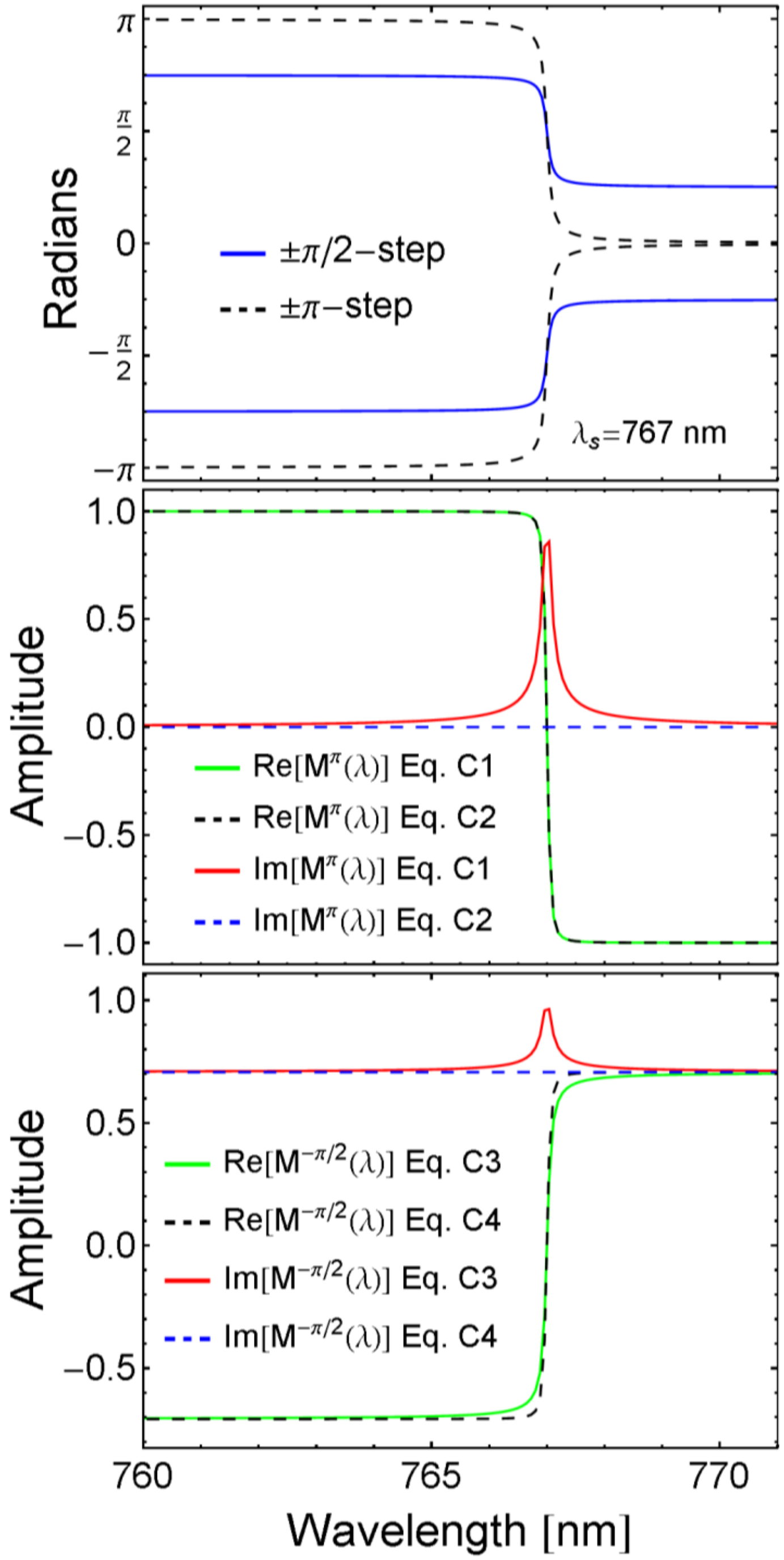

FIG. 7: (Color online) (a) Four different spectral phase profiles ($\pm\pi, \pm\pi/2$) for $\omega_s = 766\text{nm}$ and $\tau s = 2800\ \text{GHz}^{-1}$. (b) The real and imaginary parts of $M_{\pm\pi}(\omega)$, using Eq. C1 (solid) and Eq. C2 (dashed). (c) The real and imaginary parts of $M_{-\pi/2}(\omega)$, Eq. C3 (solid) and Eq. C4 (dashed).

where $\omega_s$ is the phase-step transition point and $\tau_s$ is the width parameter. Mathematically, a $+\pi$ and $-\pi$-step in Eq. C1 differ only near $\omega_s$, where it transitions positively or negatively. Near the transition region, about $\omega_s$, the argument of Eq. C1 resembles a positive or negative sloped line corresponding to a $+\pi$ and $-\pi$-step, respectively.

The real and imaginary parts of Eq. C1 are plotted in Fig. 7b. It can be seen in Fig. 7b (thin-red-solid line) that $\text{Im}\left[M_{\pm\pi}\right]$ is Lorentzian shaped and non-zero only during the step transition. The width of the Lorentzian decreases as the transition width of the phase-step decreases. We define a sharp-phase-step as one with a transition width set to the resolution of the pulse shaper. This means that the pulse-shaper does not have the needed resolution to implement the Lorentzian function. In Eq. C1, we can neglect the imaginary contribution and redefine $M_{\pm\pi}(\omega)$ as

$$M_{\pm\pi}(\omega) = \frac{(\omega-\omega_s)\tau_s}{\sqrt{(\omega-\omega_s)^2\tau_s^2+1}} \tag{C2}$$

The real and imaginary parts of Eq. C2 are plotted in Fig. 7b to compare to Eq. C1. Our assumption to neglect the Lorentzian can be verified experimentally by measuring the signal for both a $+\pi$ and $-\pi$-step and if spectra are identical then neglecting the Lorentzian was justified. If there is a difference in the spectra, then the sharpness of the transition should be increased until there is no difference between the two spectra. Neglecting the Lorentzian was also considered in Ref [31], where they considered the step function, Eq. C2, as the signum function. Here we use the arctan function and set the width small.

Using our assumption of a sharp step, the definition of the $\pm\pi/2$-step can be done in a similar manor as the $\pm\pi$-step. The expression, before we apply our assumption, for $M_{\pm\pi/2}(\omega)$ reads

$$M_{\pm\pi}(\omega) = \cos[1/2(\pi - \arctan(\omega - \omega_s)\tau_s)] \pm i\sin[1/2(\pi - \arctan[(\omega - \omega_s)\tau_s]]. \tag{C3}$$

The second term in Eq. C3 is constant away from $\omega_s$ and is a Lorentzian centered at $\omega_s$. Equation C3 has a constant phase added for convenience, such that an analytical solution can be found. After neglecting the Lorentzian from the second term, we have the following expression

$$M_{\pm\pi/2}(\omega) \approx \frac{1}{\sqrt{2}}\left(\frac{(\omega-\omega_s)\tau_s}{\sqrt{(\omega-\omega_s)^2\tau_s^2+1}} + i\right) \tag{C4}$$

In Eq. C4 we used the asymptotic limit $x \gg 1$ for the $\arctan(x)$ to derive the final expression. It turns out that this expression satisfies the condition $x \ll 1$. The real and imaginary components of Eqs. C3 and C4 are plotted in Fig. 7c for comparison. The approximated step-function (Eq. C4) has a sharper step transition then the actual function (Eq. C3). Note that Eqs. C2 and C4 have the same transition width, whereas, Eqs. C1 and C3 have slightly different transition widths.